\documentclass{article}
\usepackage{amsmath,epsfig}
\usepackage{makecell}
\usepackage{multirow}
\usepackage{booktabs}
\usepackage{rotating}
\usepackage[preprint]{spconfa4}

\let\OLDthebibliography\thebibliography
\renewcommand\thebibliography[1]{
  \OLDthebibliography{#1}
  \setlength{\parskip}{0pt}
  \setlength{\itemsep}{0pt plus 0.3ex}
}

\begin{document}\sloppy

\def\x{{\mathbf x}}
\def\L{{\cal L}}

\title{SNR-based teachers-student technique for speech enhancement}
%
\name{Xiang Hao, Xiangdong Su$^{\ast}$, Zhiyu Wang, Qiang Zhang, Huali Xu and Guanglai Gao}
\address{Inner Mongolia Key Laboratory of Mongolian Information Processing Technology,\\College of Computer Science, Inner Mongolia Univeristy, Hohhot, China \\ haoxiangsnr@gmail.com, cssxd@imu.edu.cn}

\maketitle

\begin{abstract}
  It is very challenging for speech enhancement methods to achieves robust performance under both high signal-to-noise ratio (SNR) and low SNR simultaneously. In this paper, we propose a method that integrates an SNR-based teachers-student technique and time-domain U-Net to deal with this problem. Specifically, this method consists of multiple teacher models and a student model. We first train the teacher models under multiple small-range SNRs that do not coincide with each other so that they can perform speech enhancement well within the specific SNR range. Then, we choose different teacher models to supervise the training of the student model according to the SNR of the training data. Eventually, the student model can perform speech enhancement under both high SNR and low SNR. To evaluate the proposed method, we constructed a dataset with an SNR ranging from -20dB to 20dB based on the public dataset. We experimentally analyzed the effectiveness of the SNR-based teachers-student technique and compared the proposed method with several state-of-the-art methods.
\end{abstract}
\begin{keywords}
  speech enhancement, knowledge distillation, teacher-student technique, low SNR
\end{keywords}
\section{Introduction}
\label{sec:intro}

Speech enhancement is to separate clean speech from noisy speech~\cite{se_book}. It is an essential branch of speech signal processing and has been widely studied in the past few decades. It can be used in hearing aids, voice recorders, and smart speakers, as well as the front end of tasks such as speech recognition~\cite{speech_se} and speaker recognition~\cite{speaker_se}. In recent years, a large number of speech enhancement methods based on deep learning have been proposed~\cite{se_overview}~\cite{regression_approch}~\cite{crn}~\cite{wavenet}~\cite{iconip_2019}, showing stronger robustness than traditional signal-based methods. These methods can generally be divided into time-domain methods and frequency-domain methods. The time-domain methods~\cite{unetgan}~\cite{tcnn} use the neural network to directly map noisy speech waveform to clean speech waveform and usually do not require any preprocessing. The frequency-domain methods generally use short-term Fourier transform (STFT) to convert the noisy speech from the time domain to the frequency domain. They then use the neural network to map the magnitude spectrum of the noisy speech to some masking~\cite{ibm_as_goal_wang_2005} or the magnitude spectrum of the clean speech~\cite{regression_approch}. Compared with the chaotic time-domain sampling points, the magnitude spectrum contains more geometric information, which makes it easier to calculate losses and analyze frequency components. As the SNR of noisy speech decreases, the correct phase becomes more and more important for speech intelligibility and quality~\cite{important_phase}. However, since the mapping of the phase spectrum is complicated (no obvious geometric structure), the speech enhancement methods in the time domain are also widely used.

With the increasing demand for speech-related services in recent years, the scenarios that need to be addressed for speech enhancement are becoming more and more challenging. A noticeable trend is that the range of SNR of the noisy speech is significantly expanded. Typical scenarios include stations, factories, subways, and shopping malls. The increased range of SNR means that speech enhancement needs to have the ability to enhance a wide range of background noises of different intensities. When the background noise is low, the main goal of the speech enhancement is to improve the hearing sense of the speech. When the background noise is high (in extreme cases (below -10dB), noisy speech is even hard to be heard), the speech enhancement needs to enhance the speech that is difficult to be heard to a clearer speech.

To better perform speech enhancement in the above scenarios, this paper proposes a speech enhancement method based on knowledge distillation~\cite{hinton2015distilling} and time-domain U-Net. Our motivation is as follows. Speech enhancement can be viewed as a particular task of speech separation. In speech separation, Wave-U-Net~\cite{wave_u_net}, a model in the time domain, has achieved state-of-the-art performance. It can perform feature mapping directly in the time domain to avoid processing the phase. Inspired by it, we built a powerful time-domain model suitable for speech enhancement.

To enable the speech enhancement model to handle both high SNR and low SNR, there are usually two solutions. The first solution is providing a large amount of training data at each SNR and training a large-scale neural network. The second solution is to integrate multiple different models for a specific SNR range to process the noisy speech in parallel or serially. However, the disadvantage of the former is that large-scale neural networks will consume a lot of computing resources, memories, and times. The problem of the latter is that it needs to train multiple models, which is very troublesome to deploy and severely limits the application. To deal with the above problems, we introduce the knowledge distillation, which has been widely used in image recognition~\cite{object_detection_knowledge_distillation} and speech recognition~\cite{danpovey_distilling}. Knowledge distillation can extract knowledge from a large teacher model and improve the performance of a small student model. It is also called the teacher-student technique. In this paper, we extend the traditional teacher-student technique and propose an SNR-based teachers-student technique. We first build multiple teacher speech enhancement models and train them independently with the datasets of small SNR range. Then we build a student model. To make the student model have the ability to handle both high SNR and low SNR, we will use different teacher models to guide its training according to the SNR of the training data.

To evaluate the proposed method, we construct a challenging speech enhancement dataset that covers a wide range of SNR (-20dB to 20dB). We experimentally analyzed the effectiveness of the SNR-based teachers-student technique and compared the proposed method with several state-of-the-art (SOTA) methods.

\subsection{Method}

\subsection{SNR-based teachers-student technique}

\begin{figure}
   \centering
   \includegraphics[width=\linewidth]{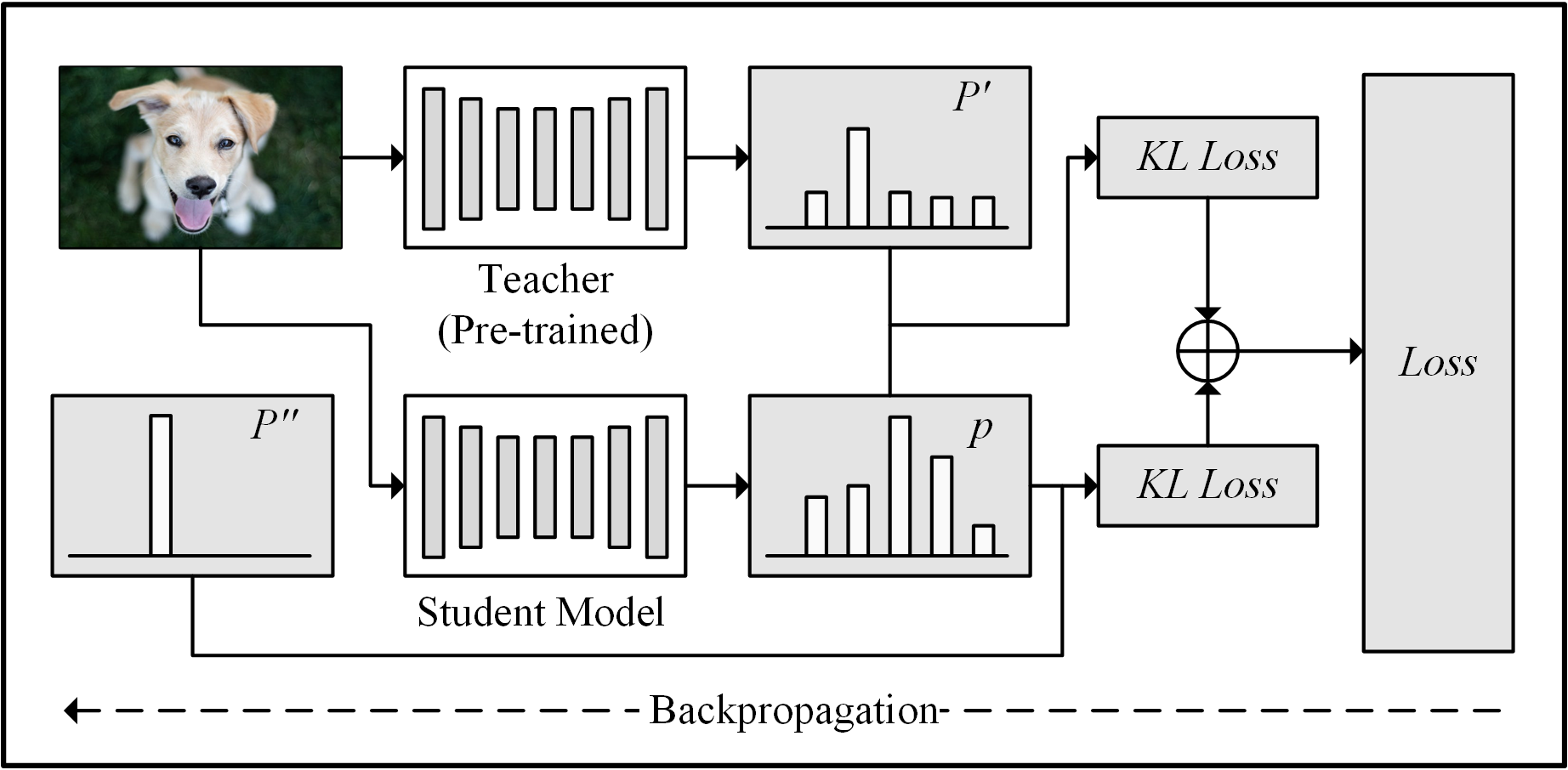}
   \caption{An illustration of the standard knowledge distillation.}
   \label{fig:teacher_and_student}
\end{figure}

\begin{figure}
   \centering
   \includegraphics[width=\linewidth]{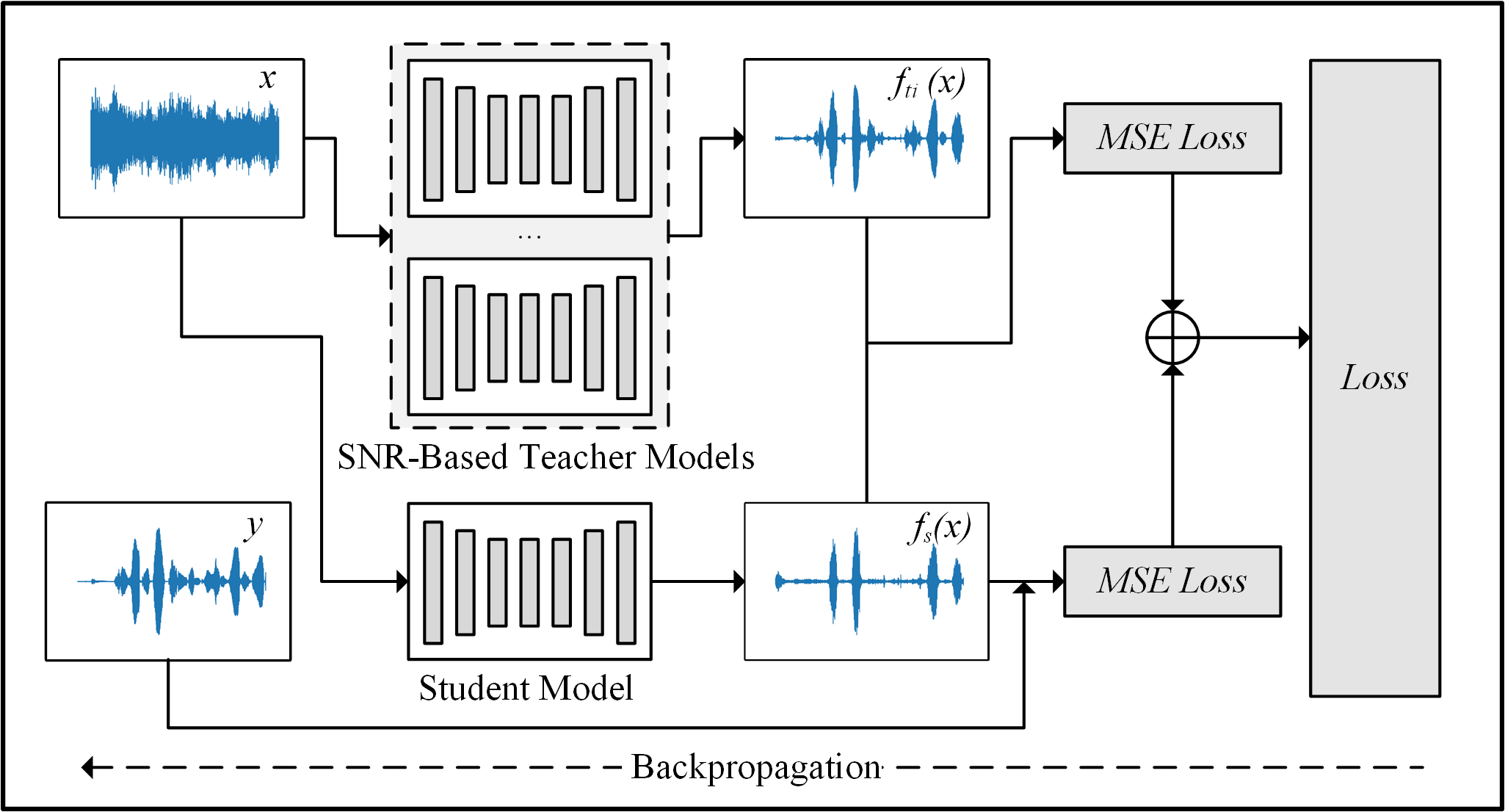}
   \caption{An illustration of the SNR-based teachers-student technique.}
   \label{fig:teachers_and_student}
 \end{figure}

We show the standard knowledge distillation in Figure~\ref{fig:teacher_and_student}. The idea behind knowledge distillation is that the soft probability output $p'$ of a pre-trained teacher model contains more information than the correct class label $p''$. If the teacher model gives higher probabilities for specific categories, then this indicates that the categories of the input image should be near these categories. Knowledge distillation forces the student model to minimize the difference between its probability output $p$ and the teacher's probability output $p'$ to extract the additional knowledge that the teacher model obtained when calculating the correct probability. We usually use knowledge distillation to distill the information learned by large networks and pass it to networks with small parameters and weak learning capabilities.
In this paper, we extend the standard knowledge distillation. The workflow are shown in Figure~\ref{fig:teachers_and_student}. First, we train multiple teacher models who are proficient in speech enhancement at a specific small SNR range. Then we use the teacher models to guide the training of the student model so that the student model can perform speech enhancement under both high SNR and low SNR. Below we describe the entire process.

\textbf{Train the teacher models.} We use $ f_T = \{ f_{t_1}, f_{t_2}, ... \}$ to represent the set of teacher models. We use $f_{t_i}$ represents the $i$-th teacher model. The dataset of each teacher model covers only a small range of SNRs, and the range of SNRs covered by each teacher's dataset does not overlap each other. After sufficient training, we get multiple teacher models that are good at dealing with the SNR covered by their dataset. We fixed the weights of the teacher models.

\textbf{Forward calculation.} We use $x$ to represent a noisy speech, and $y$ to represent the clean speech corresponding to the $x$. We input the $x$ to the SNR-based teacher models. According to the SNR of $x$, the training algorithm will select a teacher model $f_{t_i}$ to obtain the enhanced speech $f_{t_i} (x)$. We also input $x$ to the student model $f_s$ to get the enhanced speech $f_s (x)$.

\textbf{Distill knowledge from the teacher model}. We use a $L_2$ loss function to minimize the difference between the output of the student model and the teacher model. This loss can extract additional knowledge from the teacher model during the student training process, and strengthen the student model's ability to enhance speech at a specific SNR. We still need to preserve the loss between the enhanced speech of the student model $f_s (x)$ and the clean speech $y$. The overall loss $L$ is as follows:
\begin{eqnarray}
   L = \alpha \frac{1}{2} {|| f_s (x) - f_{t_i} (x) ||}^2  + (1 - \alpha)  \frac{1}{2} {|| f_s (x) - y||}^2
   \label{eq:loss}
\end{eqnarray}
where $\alpha$ is the weight of the knowledge obtained from the teachers, which is determined based on the validation dataset.

\begin{figure*}[t]
  \centerline{\includegraphics[width=\textwidth]{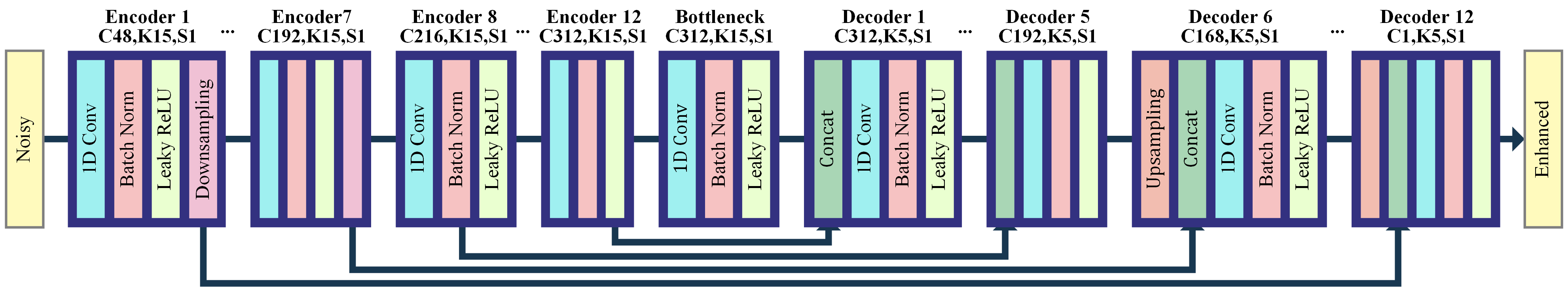}}
  \caption{The network structure of the time-domain U-Net model. $C$ represents the number of convolution kernels, $K$ represents the size of the convolution kernels, and $S$ represents the stride. All convolutional layers of the model use same-padding.}
  \label{fig:basic_network}
\end{figure*}

\subsection{Model structure}

As the SNR of noisy speech decreases, the correct phase becomes more and more critical for speech enhancement. However, since the phase spectrum mapping is complicated, we conduct speech enhancement in the time domain.
Inspired by Wave-U-Net~\cite{wave_u_net}, we designed a time-domain speech enhancement model based on one-dimensional (1D) convolutional neural networks. The teacher models and student model both take such the network structure.
The input of the model is a fixed-length noisy speech signal, and the output is an enhanced speech signal.
The model contains three parts: Encoder, Bottleneck, and Decoder, and its structure is shown in Figure~\ref{fig:basic_network}.
Below we describe it in detail.

The Encoder part consists of 12 consecutive encoder blocks.
The first seven encoder blocks use the same structure: ``1D convolutional layer + batch normalization + leaky ReLU + downsampling layer".
1D convolutional layers can be used for feature extraction and integration. The parameters of the convolutional layers are listed above each encoder block.
These seven encoder blocks all contain a downsampling layer, which can select one sampling point from two adjacent sampling points as output.
After their processing, the size of the input signal in the time dimension will gradually become smaller.
The structure of the remaining five encoder blocks in the Encoder is similar to that of the previous seven encoder blocks, but they do not contain a downsampling layer.
For the time-domain model, since the first convolution layer of the model plays a critical role~\cite{sincnet}, we set the number of convolution kernels in the first convolution layer to a larger value (48), so that the model can better extract the features of the speech waveform.
The number of convolution kernels in the remaining convolutional layers increases with the step size of 24 as the number of layers of the network increases.
The structure of the Bottleneck section is the same as that of the Encoder.

The Decoder part contains 12 consecutive decoder blocks.
The beginning of each decoder block contains the concatenation operation, which is used to concatenate the underlying features passed through the skip-connection.
In order to maintain symmetry, the last seven decoder blocks in the decoder part all include the upsampling layer.
The upsampling layer doubles the size of the feature map in the time dimension by linear interpolation, and finally restores the size corresponding to the input speech waveform.
The remaining structure of the decoder blocks are similar to the encoder blocks.

\begin{table*}[!t]
  \centering
  \scriptsize
  \caption{The PESQ and STOI of the proposed method at different SNRs and different noises.}
  \renewcommand\arraystretch{1}
  \setlength{\tabcolsep}{1.2mm}{
     \begin{tabular}{cccccccccccccccccccc}
        \toprule
        \multirow{3}[0]{*}{Noise}           & \multirow{3}[0]{*}{Target} & \multicolumn{9}{c}{PESQ} & \multicolumn{9}{c}{STOI}                                                                                                                                                                           \\
        \cmidrule(r){3-11}\cmidrule(lr){12-20}
                                            &                            & \multicolumn{5}{c}{Seen} & \multicolumn{4}{c}{Unseen} & \multicolumn{5}{c}{Seen} & \multicolumn{4}{c}{Unseen}                                                                                                                 \\
        \cmidrule(r){3-7}\cmidrule(r){8-11}\cmidrule(lr){12-16}\cmidrule(r){17-20}
                                            &                            & -20dB                    & -10dB                      & 0dB                      & 10dB                       & 20dB  & -15dB & -5dB  & 5dB   & 15dB  & -20dB & -10dB & 0dB   & 10dB  & 20dB  & -15dB & -5dB  & 5dB   & 15dB  \\
        \midrule
        \multirow{2}[0]{*}{Babble}          & Noisy                      & 0.973                    & 1.025                      & 1.438                    & 2.215                      & 2.900 & 0.703 & 1.079 & 1.823 & 2.573 & 0.326 & 0.443 & 0.641 & 0.815 & 0.912 & 0.362 & 0.547 & 0.740 & 0.871 \\
                                            & Enhanced                   & 1.410                    & 1.495                      & 2.258                    & 2.816                      & 3.227 & 1.084 & 1.783 & 2.577 & 3.018 & 0.336 & 0.552 & 0.808 & 0.901 & 0.941 & 0.377 & 0.671 & 0.861 & 0.923 \\
        \midrule
        Destroyer                           & Noisy                      & 0.924                    & 0.886                      & 1.326                    & 2.087                      & 2.777 & 0.783 & 1.050 & 1.716 & 2.442 & 0.361 & 0.462 & 0.646 & 0.839 & 0.937 & 0.399 & 0.548 & 0.750 & 0.894 \\
        engine                              & Enhanced                   & 1.230                    & 1.727                      & 2.400                    & 2.790                      & 3.143 & 1.315 & 2.083 & 2.574 & 2.966 & 0.423 & 0.637 & 0.831 & 0.899 & 0.941 & 0.482 & 0.739 & 0.868 & 0.923 \\
        \midrule
        Destroyer                           & Noisy                      & 0.411                    & 0.844                      & 1.525                    & 2.285                      & 2.997 & 0.432 & 1.077 & 1.926 & 2.640 & 0.355 & 0.472 & 0.669 & 0.817 & 0.908 & 0.400 & 0.572 & 0.753 & 0.868 \\
        ops                                 & Enhanced                   & 1.140                    & 1.762                      & 2.471                    & 2.941                      & 3.342 & 1.461 & 2.169 & 2.727 & 3.130 & 0.449 & 0.644 & 0.840 & 0.904 & 0.941 & 0.498 & 0.754 & 0.872 & 0.928 \\
        \midrule
        \multirow{2}[0]{*}{Factory floor 1} & Noisy                      & 0.855                    & 0.831                      & 1.338                    & 2.137                      & 2.894 & 0.906 & 0.989 & 1.744 & 2.516 & 0.322 & 0.437 & 0.624 & 0.817 & 0.920 & 0.373 & 0.525 & 0.727 & 0.874 \\
                                            & Enhanced                   & 1.070                    & 1.631                      & 2.322                    & 2.811                      & 3.191 & 1.296 & 1.993 & 2.538 & 2.975 & 0.368 & 0.577 & 0.803 & 0.898 & 0.941 & 0.415 & 0.708 & 0.857 & 0.921 \\
        \midrule
        Speech shaped                       & Noisy                      & 0.640                    & 0.547                      & 1.307                    & 2.116                      & 2.850 & 0.424 & 0.921 & 1.724 & 2.490 & 0.349 & 0.443 & 0.628 & 0.817 & 0.922 & 0.384 & 0.525 & 0.733 & 0.877 \\
        noise                               & Enhanced                   & 0.995                    & 1.588                      & 2.356                    & 2.820                      & 3.152 & 1.232 & 1.989 & 2.591 & 2.984 & 0.444 & 0.553 & 0.804 & 0.896 & 0.943 & 0.460 & 0.700 & 0.860 & 0.922 \\
        \midrule
        Pedestrian                          & Noisy                      & 0.924                    & 0.774                      & 1.407                    & 2.178                      & 2.902 & 0.616 & 1.061 & 1.813 & 2.518 & 0.333 & 0.461 & 0.646 & 0.815 & 0.909 & 0.390 & 0.550 & 0.737 & 0.873 \\
        area                                & Enhanced                   & 1.185                    & 1.370                      & 2.222                    & 2.802                      & 3.235 & 1.244 & 1.803 & 2.534 & 2.965 & 0.341 & 0.523 & 0.796 & 0.898 & 0.937 & 0.407 & 0.681 & 0.856 & 0.924 \\
        \midrule
        \multirow{2}[0]{*}{On the bus}      & Noisy                      & 0.636                    & 1.377                      & 2.038                    & 2.906                      & 3.608 & 0.859 & 1.753 & 2.474 & 3.198 & 0.513 & 0.703 & 0.784 & 0.860 & 0.907 & 0.610 & 0.751 & 0.832 & 0.885 \\
                                            & Enhanced                   & 1.242                    & 2.277                      & 2.746                    & 3.287                      & 3.728 & 1.835 & 2.569 & 3.039 & 3.457 & 0.517 & 0.785 & 0.874 & 0.910 & 0.938 & 0.643 & 0.839 & 0.892 & 0.924 \\
        \midrule
        \multirow{2}[0]{*}{Cafe}            & Noisy                      & 0.665                    & 0.911                      & 1.645                    & 2.404                      & 3.234 & 0.704 & 1.211 & 2.049 & 2.775 & 0.344 & 0.495 & 0.691 & 0.853 & 0.934 & 0.422 & 0.577 & 0.783 & 0.894 \\
                                            & Enhanced                   & 0.981                    & 1.584                      & 2.378                    & 2.954                      & 3.445 & 1.185 & 1.934 & 2.721 & 3.140 & 0.357 & 0.619 & 0.819 & 0.909 & 0.948 & 0.477 & 0.722 & 0.875 & 0.929 \\
        \midrule
        \multirow{2}[0]{*}{Street}          & Noisy                      & 0.806                    & 0.776                      & 1.715                    & 2.453                      & 3.139 & 0.492 & 1.205 & 2.106 & 2.799 & 0.399 & 0.545 & 0.726 & 0.841 & 0.919 & 0.508 & 0.634 & 0.800 & 0.878 \\
                                            & Enhanced                   & 1.286                    & 1.844                      & 2.589                    & 3.044                      & 3.489 & 1.591 & 2.179 & 2.851 & 3.217 & 0.402 & 0.675 & 0.857 & 0.908 & 0.941 & 0.554 & 0.778 & 0.888 & 0.927 \\
        \bottomrule
     \end{tabular}%
  }
  \label{tab:model_performance}%
\end{table*}%

\section{Experiment}
\subsection{Dataset}

We use public datasets to evaluate the proposed method.

\textbf{For SNR-based teacher models:} In this paper, we set up four teacher models. In order to train the teacher models for different SNR ranges, we randomly select 950 clean speeches from the TIMIT~\cite{timit} training dataset, and mixed speech shaped noise, babble, destroy engine, destroyer ops and factory floor 1 (from NoiseX-92 dataset~\cite{noisex92}) under different SNRs to generate the noisy speech dataset. The SNRs of the four teacher models are \{-20, -17, -13, -11\}dB, \{-10, -7, -3, 1\}dB, \{0, 3, 7, 9\}dB, and \{10, 13, 17, 20\}dB. In the end, we generated 19,000 noisy speeches for each teacher model, of which 18,000 speeches were used as the training dataset and the rest were used as the validation dataset.

\textbf{For student model:} We randomly selected 950 speeches from the TIMIT training dataset and mixed them with the five noises that appeared in the training dataset of the teacher models at \{-20, -10, 0, 10, 20\}dB. This produces 23,750 noisy speeches. Among them, 22,000 were used as the training dataset, and 1,750 were used the validation dataset.
To evaluate the student model, we randomly selected 100 speeches from the TIMIT test dataset, mixing with five noises that appeared in the above two datasets and four noises from the CHiME-4~\cite{chime4} dataset (pedestrian area noise, on the bus noise, cafe noise, and street noise) at \{-20, -15, -10, -5, 0, 5, 10, 15, 20\}dB. The resulting 8,100 noisy speeches act as the test dataset. It is worth mentioning that the test dataset of the student model contains noises, SNRs, and speakers that have not appeared in the training dataset, which makes speech enhancement very challenging.

We use PESQ~\cite{pesq} and STOI~\cite{stoi} to measure speech quality and intelligibility, respectively.

\subsection{Implementation and training details}

The sampling rate of all speeches is 16,000Hz.
The input length of all models is fixed.
Before the beginning of each training epoch, we will select consecutive 16,384 sampling points (1.024 seconds) from the random position of noisy speech as the input of the model.
Except for the learning rate, other experimental parameters are the same for all models.
We use Adam optimizer~\cite{adam} (decay rates $\beta_{1} = 0.9$, $\beta_{2} = 0.999$) and set the batch size to 16. we set $\alpha$ of leaky ReLU to 0.1.
We set the learning rate of the teacher models to a small constant of 0.0002.
We set the initial learning rate of the student model to 0.002, which is reduced twice by every 300 epochs until the validation loss does not decrease.
According to the validation dataset, we set $\alpha$ in Equation~\ref{eq:loss} to $0.5$.

\subsection{Baseline methods}

We compared with the three methods, which are Wavenet-denoising~\cite{wavenet}, PSM-BiLSTM~\cite{phase_sensitive_bilstm}, and CRN~\cite{crn}.
The first one is to mapping features on the time domain, and the last two are mapping features on the frequency domain.
\begin{itemize}
   \item Wavenet-denoising: It is an end-to-end method for speech denoising based on wavenet~\cite{wavenet_sn}. It retains wavenet’s powerful acoustic modeling capabilities, while significantly reducing time complexity by eliminating its autoregressive nature. We used its official implementation.
   \item PSM-BiLSTM: It is a bidirectional LSTM network for speech enhancement, using a phase-sensitive spectrum approximation (PSA) cost function. We implemented it and used the same hyperparameters as that in the paper.
   \item CRN: It contains a convolutional encoder-decoder and a LSTM bottleneck. We implemented it and used the same hyperparameters as that in the original paper.
\end{itemize}

\section{Result}

\subsection{Performance of the proposed method}

Table~\ref{tab:model_performance} presents the noisy speech and the enhanced speech from -20dB to 20dB in terms of PESQ and STOI. The ``Noisy” lines in the table represent the noisy speech, and the ``Enhanced” lines represent the enhanced speech using the proposed method. The ``Seen” columns mean the SNRs exist in the training dataset of the student model while the “Unseen” columns indicate the SNRs that do not exist in the training dataset of the student model.

When comparing the noisy and enhanced speech, we can found that the PESQ and STOI are improved after speech enhancement with the proposed method even some noises do not appear in the training dataset (the last four noises).
It is also clear that our method improves the PESQ and STOI for both “Seen” SNR and “Unseen” SNRs.
The SNRs lower than -5dB are generally considered as extremely low SNRs, and the speech enhancement under such conditions is very challenging. However, our method works well, even if the range of the SNR of the dataset is so wide (-20dB to 20dB).
The average improvement of the PESQ and STOI is 38.71\% and 12.73\%, respectively.
This improvement indicates that the proposed method is effective.


\subsection{Effectiveness of SNR-based teachers}

To demonstrate the effectiveness of the SNR-based teacher models, Table~\ref{tab:compare_student} lists the performance of the four SNR-based teacher models (T1, T2, T3, T4), the student model trained without teacher supervision (S1), and the proposed method (S2).
As shown in Table~\ref{tab:compare_student}, we only test the teacher models on the trained small range of SNRs.
For clarity, the PESQ and STOI of the noisy speech are also listed in the table.

Comparing with the noisy speech, there are notable improvements on the PESQ and STOI for the teacher models on their corresponding SNRs. This suggests that the teacher models can perform well on a small range of SNRs. The S1 and the S2 use the same training dataset and network structure. Compared to all the above models, the S1 performs the worst no matter which SNR. We guess that this is because the SNR range of the dataset is too large. With the learning ability of the S1, it is very difficult to consider all the SNRs from low to high.

The PESQ and STOI of the proposed method S2 approximate those of the teacher models at the corresponding SNR, and are far better than the S1 at all SNRs. These results indicate that the proposed SNR-based teachers-student technique can help the student model improve its ability to handle high SNR and low SNR at the same time.



\begin{figure}
  \begin{minipage}[h]{0.48\linewidth}
     \centering
     \centerline{\includegraphics[width=1\linewidth]{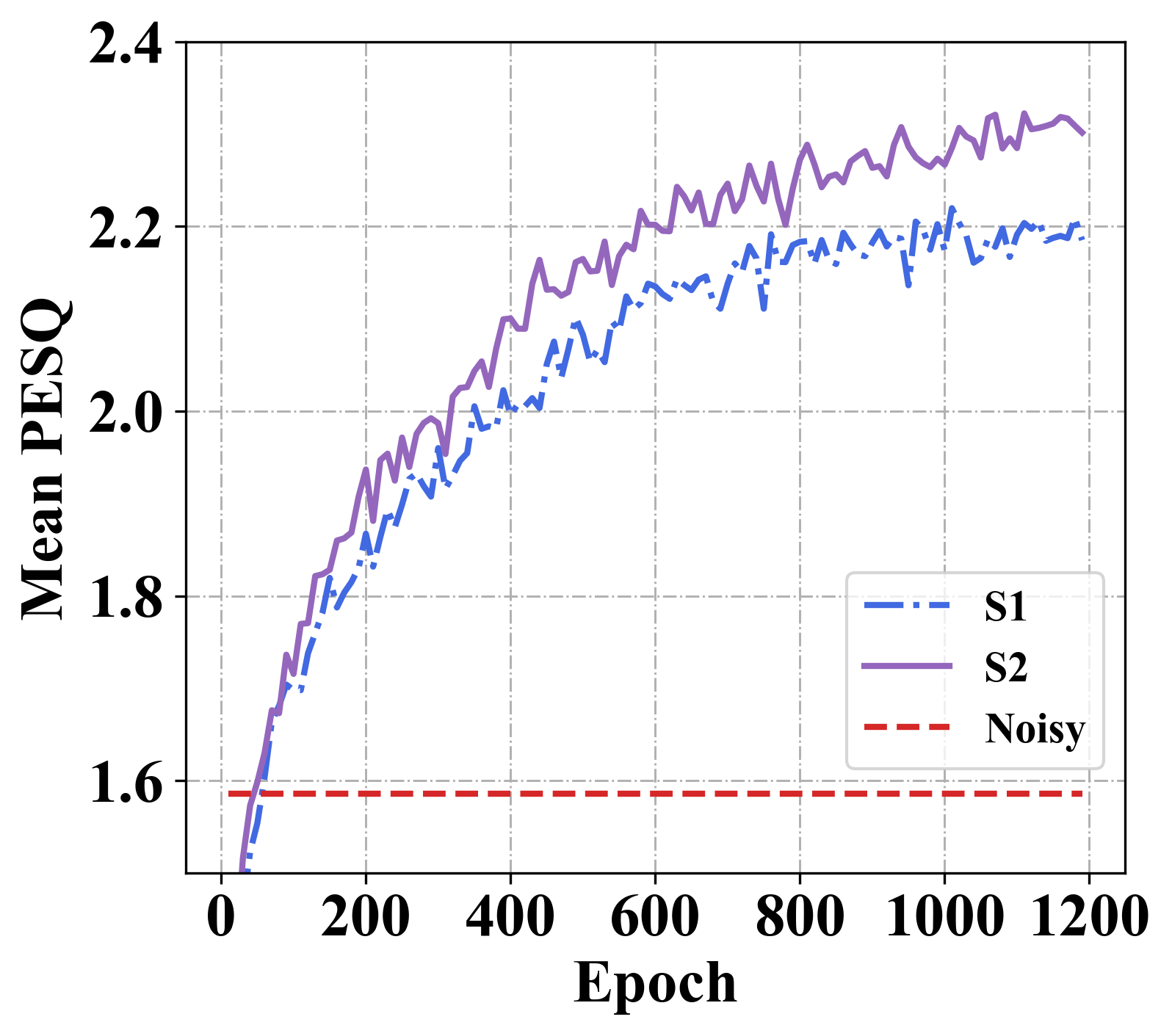}}
  \end{minipage}
  \hfill
  \begin{minipage}[h]{0.49\linewidth}
     \centering
     \centerline{\includegraphics[width=1\linewidth]{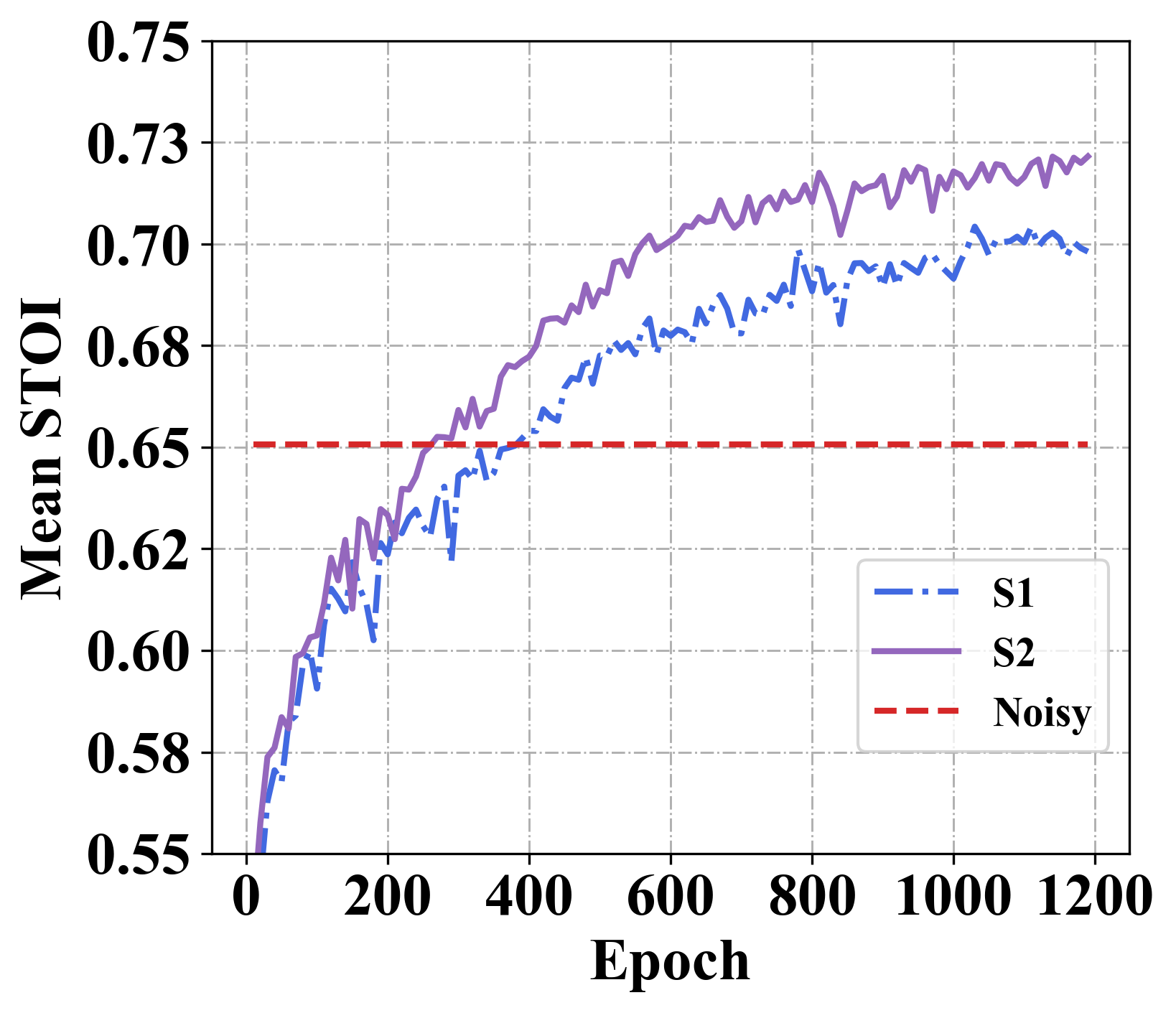}}
  \end{minipage}
  \caption{The effectiveness of the SNR-based teachers-student technique.}
  \label{fig:validation_metrics}
\end{figure}

We calculated the PESQ and STOI on the validation dataset for every ten epochs during the training process of the S1 and S2.
The growth curves are shown in Figure~\ref{fig:validation_metrics}.
The horizontal axis indicates the epoch of training, and the vertical axis is the average metric.
We can clearly notice that from about the 200th epoch, the PESQ and STOI of S2 exceed the PESQ and STOI of S1.
In the subsequent training process, the performance gap between the two models is still increasing.

\begin{table}
   \centering
   \scriptsize
   \caption{An Illustration of average performance on noisy speech (Noisy), the SNR-based teacher models (T1, T2, T3, and T4), the student without teacher (S1), and the student with SNR-based teachers (S2).}
   \renewcommand\arraystretch{1}
   \setlength{\tabcolsep}{1.2mm}{
      \begin{tabular}{ccccccccccc}
         \toprule
                                                            &       & -20dB          & -15dB          & -10dB          & -5dB           & 0dB            & 5dB            & 10dB           & 15dB           & 20dB           \\
         \midrule
         \multirow{7}[2]{*}{\begin{sideways}PESQ\end{sideways}}  & Noisy & 0.761          & 0.660          & 0.892          & 1.155          & 1.531          & 1.936          & 2.314          & 2.666          & 3.039          \\
 
         \cmidrule(lr){2-11}
                                                            & T1    & 1.132          & \textbf{1.400} & \textbf{1.720} & -              & -              & -              & -              & -              & -              \\
                                                            & T2    & -              & -              & 1.719          & \textbf{2.121} & \textbf{2.461} & -              & -              & -              & -              \\
                                                            & T3    & -              & -              & -              & -              & 2.387          & \textbf{2.761} & \textbf{3.109} & -              & -              \\
                                                            & T4    & -              & -              & -              & -              & -              & -              & 2.891          & \textbf{3.228} & \textbf{3.561} \\
         \cmidrule(lr){2-11}
                                                            & S1    & 1.018          & 1.177          & 1.581          & 1.957          & 2.324          & 2.642          & 2.885          & 3.016          & 3.257          \\
 
                                                            & S2    & \textbf{1.172} & 1.360          & 1.700          & 2.057          & 2.417          & 2.686          & 2.921          & 3.098          & 3.332          \\
         \midrule
         \multirow{7}[2]{*}{\begin{sideways}STOI\end{sideways}} & Noisy & 0.367          & 0.428          & 0.497          & 0.582          & 0.674          & 0.762          & 0.831          & 0.879          & 0.919          \\

         \cmidrule(lr){2-11}
                                                            & T1    & \textbf{0.414} & \textbf{0.498} & 0.626          & -              & -              & -              & -              & -              & -              \\
                                                            & T2    & -              & -              & \textbf{0.639} & \textbf{0.755} & \textbf{0.833} & -              & -              & -              & -              \\
                                                            & T3    & -              & -              & -              & -              & 0.827          & \textbf{0.888} & 0.918          & -              & -              \\
                                                            & T4    & -              & -              & -              & -              & -              & -              & \textbf{0.964} & \textbf{0.982} & \textbf{0.987} \\
         \cmidrule(lr){2-11}
                                                            & S1    & 0.312          & 0.441          & 0.586          & 0.693          & 0.810          & 0.865          & 0.901          & 0.915          & 0.935          \\

         \cmidrule(lr){2-11}
                                                            & S2    & 0.404          & 0.480          & 0.619          & 0.733          & 0.826          & 0.876          & 0.903          & 0.925          & 0.941          \\

         \bottomrule
      \end{tabular}%
   }
   \label{tab:compare_student}%
 \end{table}%

\subsection{Comparison of our method with the baselines}

Table~\ref{tab:compare_with} reports the performances of the proposed method and the baseline methods on the test dataset (-20dB to 20dB).
We can easily notice that the proposed method (S2) achieves the best performance, whether it is PESQ or STOI.
Compared to the most robust method CRN in the baseline methods, we still improved 0.143 on PESQ and 0.021 on STOI.
In the scenario that contains both high SNR and low SNR noisy speeches, our proposed method based on SNR-based teachers-student technique and time-domain U-Net is undoubtedly more advantageous.

\begin{table}[h]
  \centering
  \footnotesize
  \caption{Comparison of the proposed method and the baselines.} \label{tab:compare_with}
  \renewcommand\arraystretch{1.2}
  \setlength{\tabcolsep}{5mm}{
     \begin{tabular}{lrr}
        \toprule
        Method            & PESQ           & STOI           \\
        \midrule
        Noisy             & 1.661          & 0.660          \\
        Wavenet-denoising & 1.948          & 0.711          \\
        PSA-BiLSTM        & 2.020          & 0.719          \\
        CRN               & 2.161          & 0.723          \\
        S2                & \textbf{2.304} & \textbf{0.744} \\
        \bottomrule
     \end{tabular}
  }
\end{table}


\section{Conclusion}

Speech enhancement for both low SNR and high SNR is a very challenging task. This paper proposes a method that integrates an SNR-based teachers-student technique and time-domain U-Net to deal with this problem. The student model is trained with the supervision of multiple teacher models. Each teacher model is well trained in an SNR-based way, which means they are only responsible for speech enhancement of a small SNR range. The experimental result shows that our method achieves state-of-the-art performance and suggests the effectiveness of SNR-based knowledge distillation in speech enhancement.
The result also proves that our method is robust on ``Seen" and ``Unseen" noise and SNRs.
To our best knowledge, this is the first time that knowledge distillation is investigated in speech enhancement.

\section{ACKNOWLEDGMENTS}
This work was funded by National Natural Science Foundation of China (Grant No.61762069, 61773224), Natural Science Foundation of Inner Mongolia Autonomous Region (Grant No. 2017BS0601), and Inner Mongolia University Research and Innovation Project (Grant No. 10000-15010109).

\bibliographystyle{IEEEbib}
\bibliography{icme2020template}

\end{document}